\documentclass{PoS}

\usepackage{pstricks}
\usepackage{pst-plot}
\usepackage{epsfig}
\usepackage{amsmath}
\usepackage{dsfont}
\usepackage{graphicx}

\title{Twisted mass lattice computation of charmed mesons with focus on $D^{\ast \ast}$}

\ShortTitle{Twisted mass lattice computation of charmed mesons with focus on $D^{\ast \ast}$}

\author{\speaker{Martin Kalinowski}\\
        Goethe University Frankfurt am Main\\
        E-mail: \email{kalinowm@th.physik.uni-frankfurt.de}}

\author{Marc Wagner\\
        Goethe University Frankfurt am Main\\
        E-mail: \email{mwagner@th.physik.uni-frankfurt.de}}

\abstract{We present results of a 2+1+1 flavor twisted mass lattice QCD computation of the spectrum of $D$ mesons and $D_s$ mesons and of charmonium. Particular focus is put on the positive parity $D$ states (so-called $D^{\ast \ast}$ mesons) with quantum numbers $J^P = 0^+$, $1^+$ and $2^+$. Besides computing their masses we are also separating and classifying the two $J^P = 1^+$ states according to the angular momentum/spin of their light degrees of freedom (light quarks and gluons) $j =1/2, 3/2$.}

\FullConference{31st International Symposium on Lattice Field Theory LATTICE 2013\\
		July 29 -- August 3, 2013\\
		Mainz, Germany}

\begin{document}


\section{Introduction}

There is considerable interest in the spectrum of $D$ and $D_s$ mesons and of charmonium both theoretically and experimentally.

On the theory side first principles calculations are usually lattice QCD computations (for recent work cf.\ e.g.\ \cite{Mohler:2011ke,Namekawa:2011wt,Liu:2012ze,Dowdall:2012ab,Kalinowski:2012re,Bali:2012ua,Moir:2013ub,Kalinowski:2013wsa}). In the last couple of years a lot of progress has been made, allowing the determination of hadron masses like the aforementioned mesons with rather high precision. For example 2+1 or even 2+1+1 flavors of dynamical quarks are often used as well as small lattice spacings and improved discretizations, to keep discretization errors (in particular those associated with the heavy charm quarks) under control. Some groups have even started to determine the resonance parameters of certain mesons from the spectrum of two-particle scattering states in finite spatial volumes (cf.\ e.g.\ \cite{Mohler:2012na,Mohler:2013rwa}).

Experimentally a large number of $D$, $D_s$ and charmonium states has been measured and additional and more precise results are expected in the near future both from existing facilities and facilities currently under construction, like the PANDA experiment at FAIR. Even though these experimental results have been extremely helpful, to improve our understanding of QCD, they also brought up new and yet unanswered questions. For example the positive parity mesons $D_{s0}^*$ and $D_{s1}$ are unexpectedly light, which is at the moment not satisfactorily understood and also quite often not reproduced by lattice QCD computations or model calculations.

Moreover, performing a precise computation of certain meson masses is often the first step for many lattice projects not primarily concerned with spectroscopy. As an example one could mention the semileptonic decay of $B$ and $B^*$ mesons into positive parity $D$ mesons \cite{Bigi:2007qp} (so-called $D^{\ast \ast}$ mesons). Their masses and operator contents, which are discussed in detail in section~\ref{SEC006}, are an essential ingredient for any corresponding lattice computation.

This is mainly a status report about an ongoing lattice QCD project concerned with the computation of the spectrum of mesons with at least one charm valence quark. We present preliminary results for $D$ mesons, for $D_s$ mesons and for charmonium states with total angular momentum $J = 0, 1$ and parity $P = -, +$. Parts of this work have already been published \cite{Kalinowski:2012re,Kalinowski:2013wsa}.


\section{\label{SEC005}Simulation and analysis setup}

We use gauge link configurations generated by the European Twisted Mass Collaboration (ETMC) with the Iwasaki gauge action and $N_f = 2+1+1$ flavors of Wilson twisted mass quarks \cite{Baron:2010bv,Baron:2011sf,Baron:2010th,Baron:2010vp}. Until now we have considered three ensembles (around 1000 gauge link configurations per ensemble) with (unphysically heavy) values for the light $u/d$ quark mass corresponding to $m_\pi  \approx 285 \, \textrm{MeV}, 325 \, \textrm{MeV}, 457 \, \textrm{MeV}$ and lattice sizes $(L/a)^3 \times T/a = 32^3 \times 64, 32^3 \times 64, 24^3 \times 48$. Our results are obtained at a single lattice spacing $a \approx 0.086 \, \textrm{fm}$. Consequently, a continuum extrapolation has not yet been performed.

Meson masses are determined by computing and studying temporal correlation matrices of suitably chosen meson creation operators $\mathcal{O}_j$. At the moment we exclusively consider quark antiquark operators. The quark and the antiquark are combined in spin space via $\gamma$ matrices and in color and position space via gauge links such that the corresponding trial states $\mathcal{O}_j | \Omega \rangle$ ($| \Omega \rangle$ denotes the vacuum) are gauge invariant and have defined total angular momentum and parity (cf.\ section~\ref{SEC006} for examples of $J=1$ $D$ meson creation operators and \cite{Weber:2013eba}, in particular section~4.1, for a general discussion). Moreover, APE and Gaussian smearing is used, to optimize the overlap of the trial states $\mathcal{O}_j | \Omega \rangle$ to the low lying mesonic states of interest. We plan to discuss these operators and their structure and quantum numbers in detail in an upcoming publication. For the computation of the corresponding correlation matrices $\langle \mathcal{O}_j^\dagger(t) \mathcal{O}(0) \rangle$ we resort to the one-end trick (cf.\ e.g.\ \cite{Boucaud:2008xu}). Meson masses are then determined from plateau values of corresponding effective masses, which we obtain by solving generalized eigenvector problems (cf.\ e.g.\ \cite{Blossier:2009kd}). Disconnected diagrams appearing in charmonium correlators are currently neglected.

One of the main advantages of the Wilson twisted mass discretization is automatic $\mathcal{O}(a)$ improvement of physical observables, e.g.\ hadron masses. However, parity and isospin (in case of a non-degenerate quark doublet flavor instead of isospin) are not exact symmetries. For example positive and negative parity trial states are not anymore orthogonal, which leads to additional difficulties, when doing hadron spectroscopy: positive and negative parity states have to be determined from a single correlation matrix, which is typically twice as large compared to those studied in parity and isospin symmetric lattice discretizations.

For both the valence strange and charm quarks we use degenerate twisted mass doublets, i.e.\ a different discretization as for the corresponding sea quarks. We do this, to avoid mixing of strange and charm quarks, which inevitably takes place in a unitary setup, and which is particularly problematic for hadrons containing charm quarks \cite{Baron:2010th,Baron:2010vp}. The degenerate valence doublets allow two realizations for strange as well as for charm quarks, either with a twisted mass term $+i \mu_{s,c} \gamma_5$ or $-i \mu_{s,c} \gamma_5$. For a quark antiquark meson creation operator the sign combinations $(+,-)$ and $(-,+)$ for the quark $q$ and the antiquark $\bar{q}$ are related by symmetry, i.e.\ the corresponding correlators are identical. These correlators differ, however, from their counterparts with sign combinations $(+,+)$ and $(-,-)$, due to different discretization errors. In section~\ref{SEC001} we will show for each computed meson mass both the $(+,-) \equiv (-,+)$ and the $(+,+) \equiv (-,-)$ result. The differences are $\mathcal{O}(a^2)$, due to the aforementioned automatic $\mathcal{O}(a)$ improvement inherent to the Wilson twisted mass formulation. These mass differences give a first impression regarding the magnitude of discretization errors at our currently used lattice spacing $a \approx 0.086 \, \textrm{fm}$.

Using $(+,-) \equiv (-,+)$ correlators we have tuned the bare valence strange and charm quark masses $\mu_s$ and $\mu_c$ to reproduce the physical values of $2 m_K^2 - m_\pi^2$ and $m_D$, quantities, which strongly depend on $\mu_s$ and $\mu_c$, but which are essentially independent of the light $u/d$ quark mass.


\section{\label{SEC001}Numerical results}


\subsection{The $D$ meson, the $D_s$ meson and the charmonium spectrum}

\begin{figure}[htb]
\begin{center}
\includegraphics[width=7.5cm]{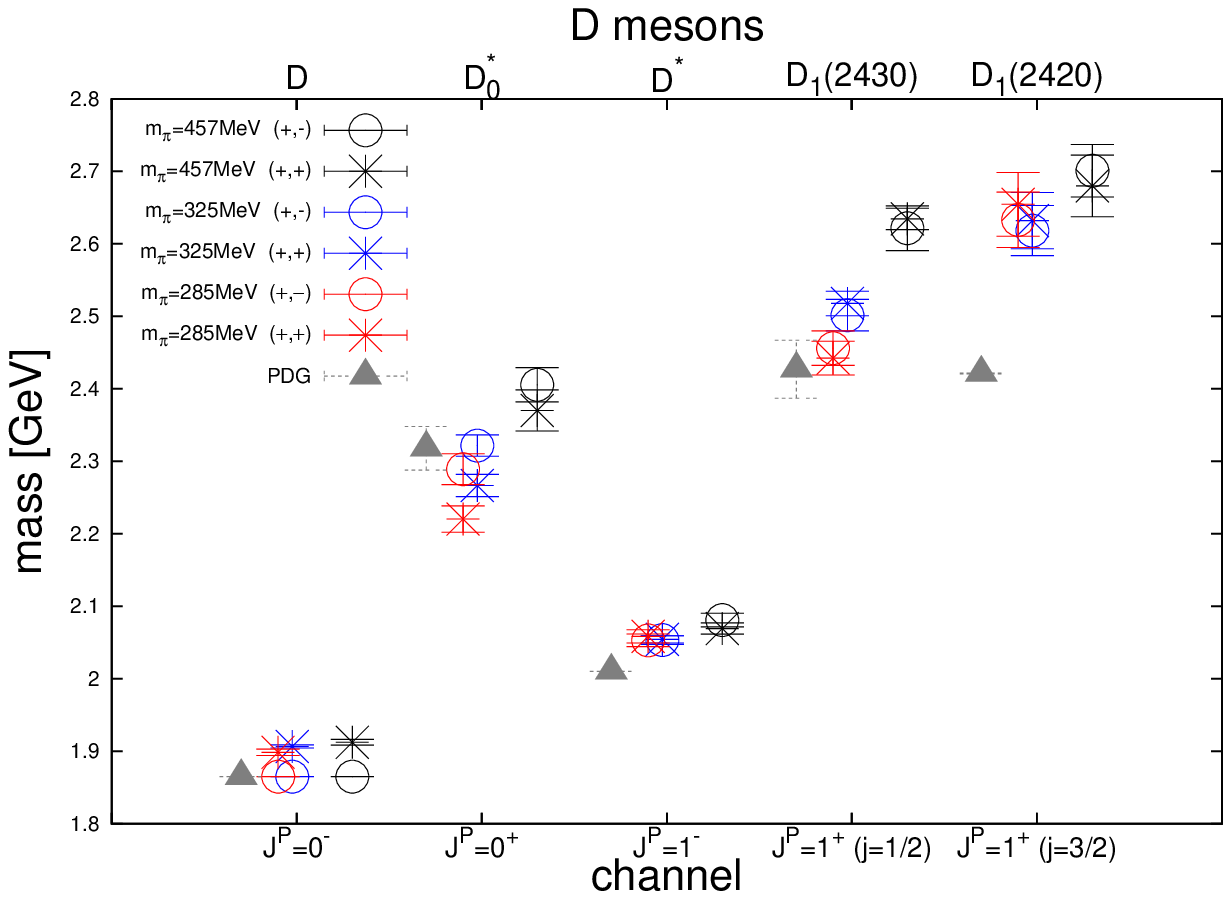} 
\includegraphics[width=7.5cm]{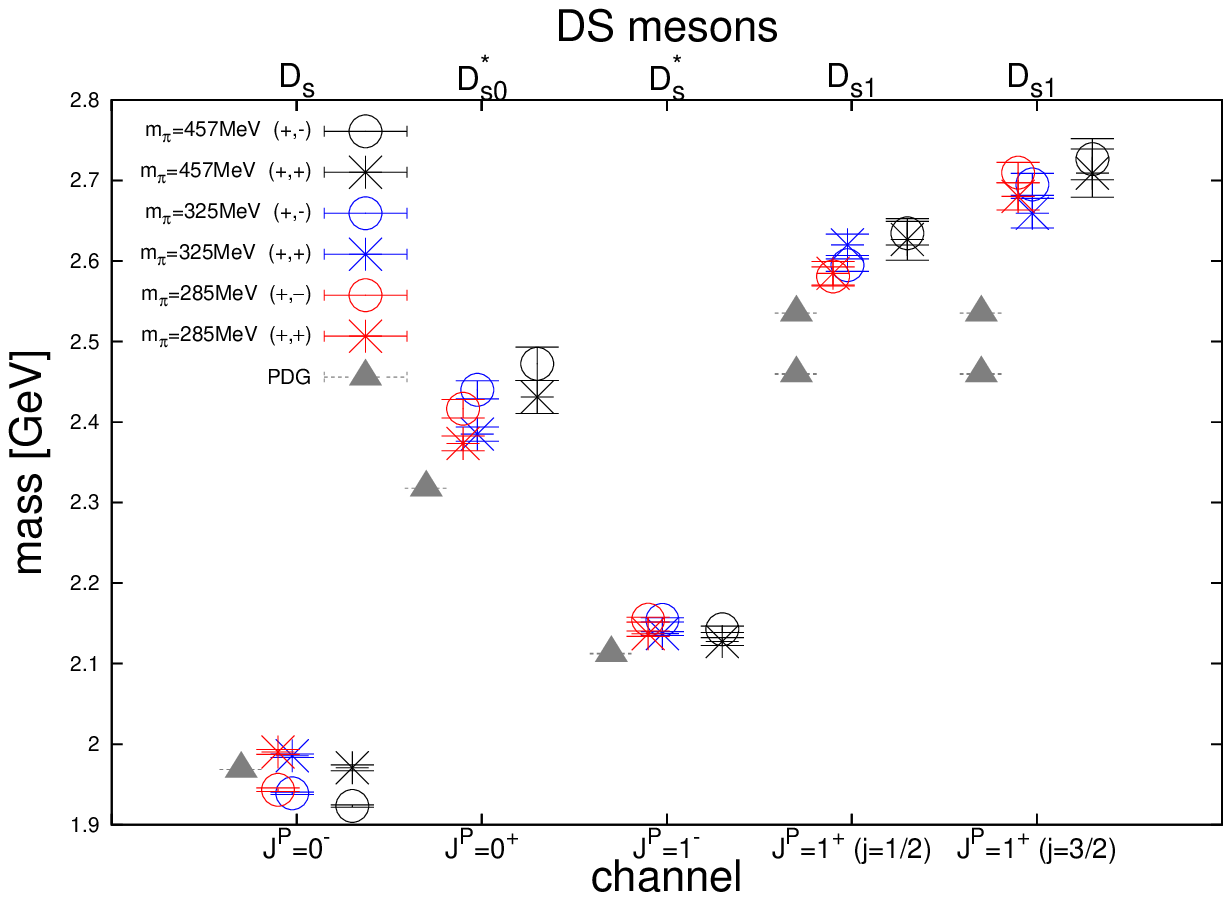} \\
\includegraphics[width=7.5cm]{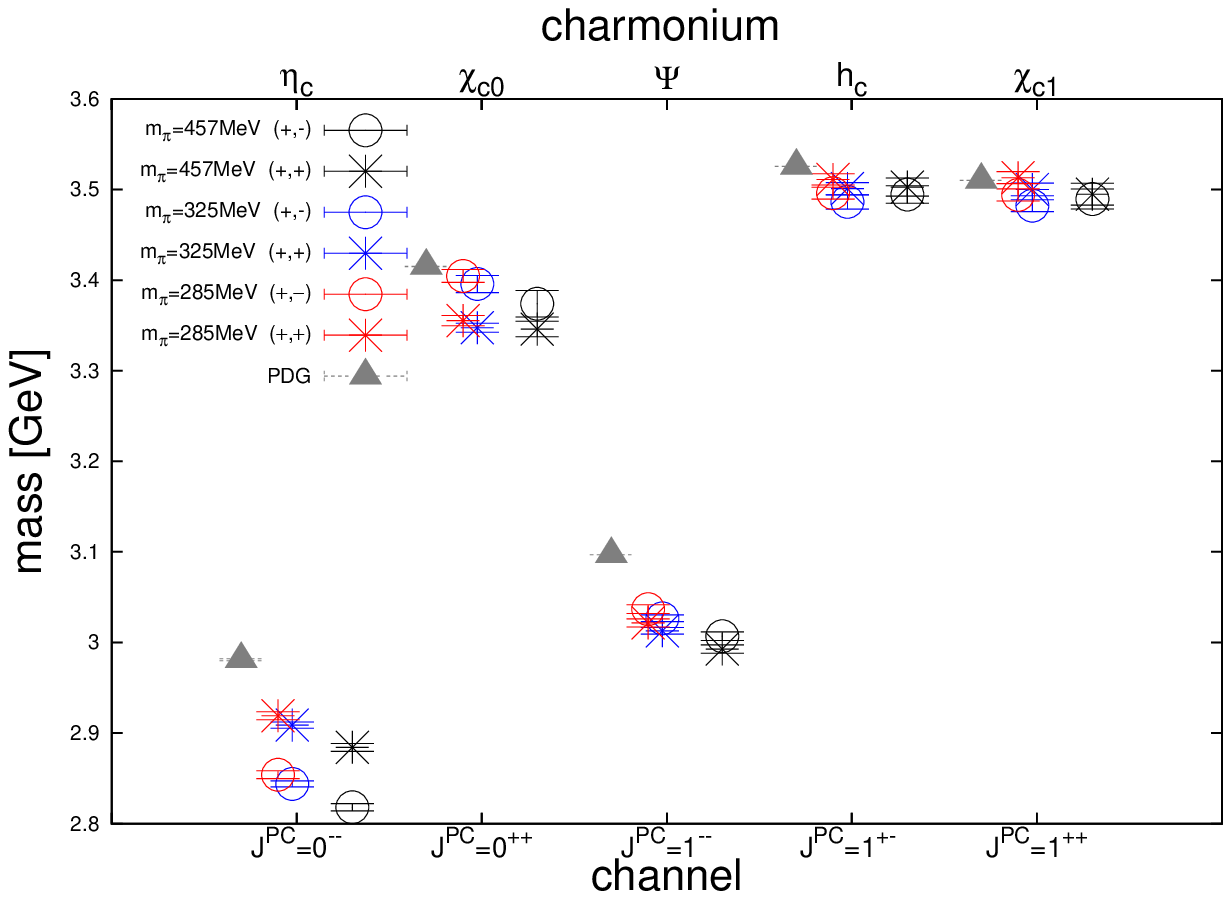}
\end{center}
\vspace{-0.5cm}
\caption{\label{FIG001}The $D$ meson, the $D_s$ meson and the charmonium spectrum for three different light quark masses corresponding to $m_\pi  \approx 285 \, \textrm{MeV}, 325 \, \textrm{MeV}, 457 \, \textrm{MeV}$ and lattice spacing $a \approx 0.086 \, \textrm{fm}$.}
\end{figure}

In Figure~\ref{FIG001} we present our results for the $D$ and $D_s$ meson spectrum. For every state we show six data points: different colors indicate the different light quark/pion masses of the used ensembles, the circles and crosses distinguish the twisted mass sign combinations $(+,-) \equiv (-,+)$ and $(+,+) \equiv (-,-)$, respectively. The horizontal separation of the data points have been chosen proportional to the corresponding squared pion masses.

While for the negative parity states lattice and experimental results agree rather well, there is a clear discrepancy in particular for the positive parity $D_s$ states $D_{s0}^*$ and $D_{s1}$. Similar findings have been reported in other lattice studies, e.g.\ \cite{Mohler:2011ke,Moir:2013ub}, and in phenomenological model calculations, e.g.\ \cite{Ebert:2009ua}. This discrepancy might be an indication that these states are not predominantly $q \bar{q}$ states, but e.g.\ rather four quark states of molecular or tetraquark type. We plan to investigate this possibility within our setup in the near future. The necessary techniques have already been developed and recently been applied to light scalar mesons \cite{Alexandrou:2012rm,Wagner:2013jda}.

In Figure~\ref{FIG001} we also present our results for the charmonium spectrum. Because of the two rather heavy valence quarks, we expect considerably larger discretization errors than for the corresponding $D$ or $D_s$ meson states. The differences between lattice and experimental results are most prominent for the negative parity charmonium states (around $5 \%$). We plan to explore in one of our next steps, whether discretization errors account for these differences by performing similar computations on ensembles with finer lattice spacings and by studying the continuum limit.


\subsection{\label{SEC006}$J^P = 1^+$ $D$ mesons: separation of the two $D_1$ states}

A challenging, but important task is the separation of the two $J=1^+$ $D$ meson states $D_1(2430)$ and $D_1(2420)$. In the limit of infinitely heavy charm quarks the broad $D_1(2430)$ state is expected to have light cloud angular momentum $j = 1/2$, while the narrow $D_1(2420)$ state should have $j = 3/2$ (cf.\ \cite{Jansen:2008si,Michael:2010aa} for a detailed discussion and computation of the static limit). Assigning corresponding approximate $j$ quantum numbers, when using charm quarks of finite mass, is e.g.\ important, when studying the decay of a $B$ or $B^*$ meson into one of the positive parity $D^{\ast \ast}$ mesons (which include the mentioned $D_1(2420)$ and $D_1(2430)$ states) in a fully dynamical setup (cf.\ e.g.\ \cite{Blossier:2009vy,Blossier:2009eg} for a recent lattice computation in the static limit and \cite{Atoui:2013sca} for first results obtained with dynamical charm quarks).

The correct identification of the $j \approx 1/2$ and the $j \approx 3/2$ state can be achieved by studying the eigenvectors obtained during the analysis of correlation matrices, i.e.\ when solving generalized eigenvector problems. After a suitable normalization of the trial states $\mathcal{O}_j | \Omega \rangle$ large eigenvector components point out the dominating meson creation operators $\mathcal{O}_j$, which, after a Clebsch-Gordan decomposition into light and heavy total angular momentum contributions, can be classified according to $j = 1/2$ or $j = 3/2$.

We use quark-antiquark meson creation operators
\begin{eqnarray}
\mathcal{O}_{\Gamma} \ \ = \ \ \sum_\mathbf{r} \, \bar{c}(\mathbf{r}) \sum_{\mathbf{n} = \pm \mathbf{e}_x,\pm \mathbf{e}_y,\pm \mathbf{e}_z} U(\mathbf{r};\mathbf{r}+\mathbf{n}) \Gamma(\mathbf{n}) u(\mathbf{r} + \mathbf{n}) ,
\end{eqnarray}
where $\bar{c}$ and $u$ are Gaussian smeared quark fields, $U(\mathbf{r};\mathbf{r}+\mathbf{n})$ is the APE smeared link connecting $\mathbf{r}$ and $\mathbf{r}+\mathbf{n}$ and $\Gamma(\mathbf{n})$ denotes suitably chosen linear combinations of products of $\gamma$ matrices and spherical harmonics realizing the desired quantum numbers $J$, $j$ and $P$. In total we consider 36 meson creation operators:
\begin{itemize}
\item for $J = 1$ and $j = 1/2$
\begin{eqnarray}
\label{EQN101} \Gamma(\mathbf{n}) \ \ = \ \ \gamma_j \gamma_5 G \quad , \quad \Gamma(\mathbf{n}) \ \ = \ \ \Big((\mathbf{n} \times \vec{\gamma})_j - \mathbf{n}_j \gamma_0 \gamma_5\Big) G ,
\end{eqnarray}

\item for $J = 1$ and $j = 3/2$
\begin{eqnarray}
\label{EQN102} \Gamma(\mathbf{n}) \ \ = \ \ \Big((\mathbf{n} \times \vec{\gamma})_j + 2 \mathbf{n}_j \gamma_0 \gamma_5\Big) G
\end{eqnarray}
\end{itemize}
with $j = 1,2,3$ and $G = 1,\gamma_0,\gamma_5,\gamma_0 \gamma_5$. Meson creation operators, which only differ in $j$, are related by symmetry. The resulting correlation functions have been averaged, to increase statistical accuracy. Meson creation operators with $G = 1, \gamma_0$ and with $G = \gamma_5, \gamma_0 \gamma_5$ correspond to $P = +$ and $P = -$, respectively. Due to twisted mass parity mixing (cf.\ section~\ref{SEC005}), these operators do not generate orthogonal trial states and, therefore, have to be included in a single $12 \times 12$ correlation matrix.

In Figure~\ref{FIG003} we show the operator content of the three lightest $J = 1$ states as a function of the temporal separation of the correlation matrix (for a detailed explanation of such plots we refer to \cite{Baron:2010th}):
\begin{figure}[htb]
\begin{center}
\includegraphics[width=7.5cm]{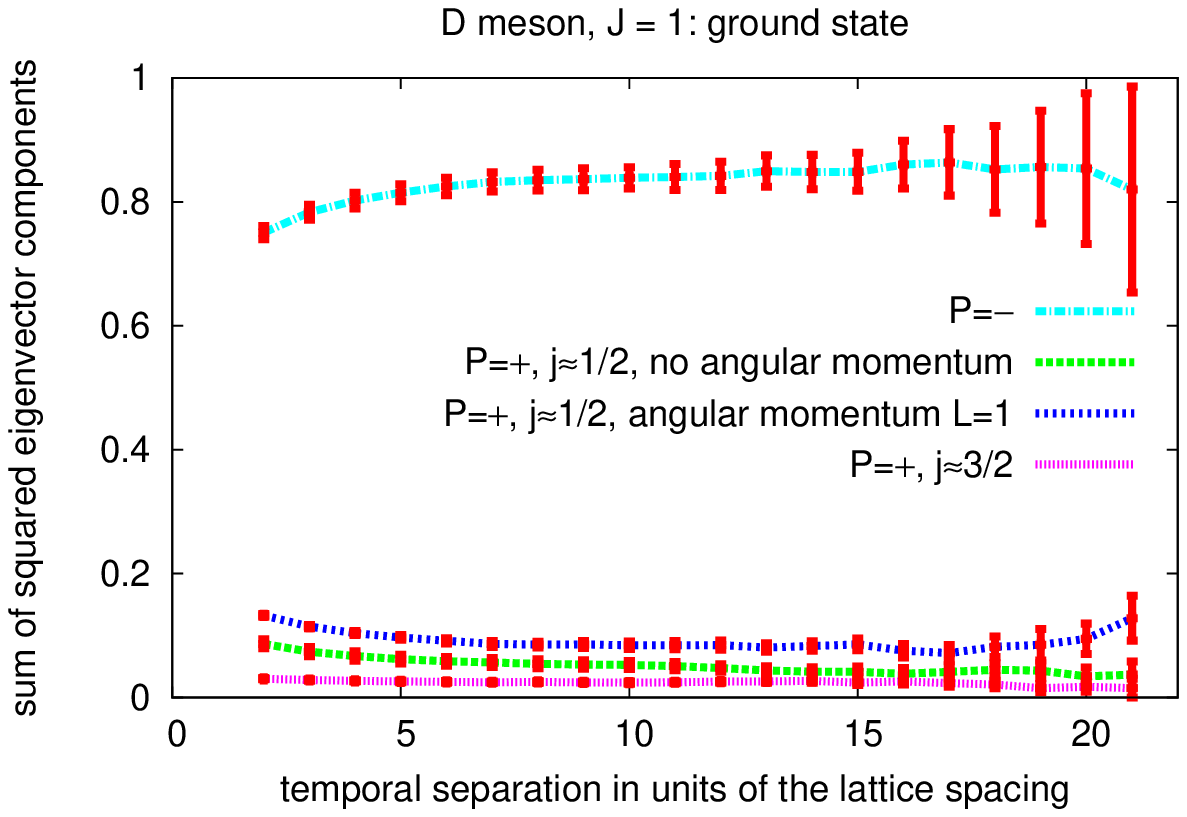} \\
\includegraphics[width=7.5cm]{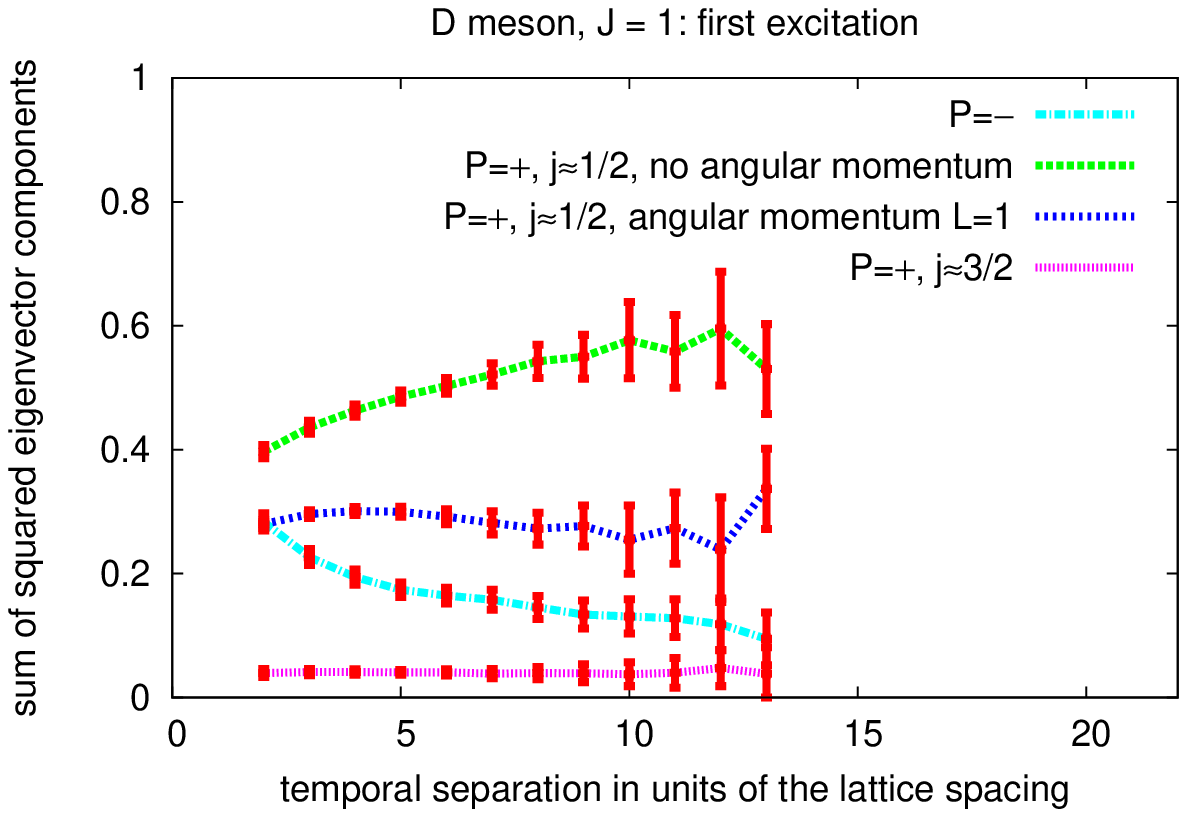} 
\includegraphics[width=7.5cm]{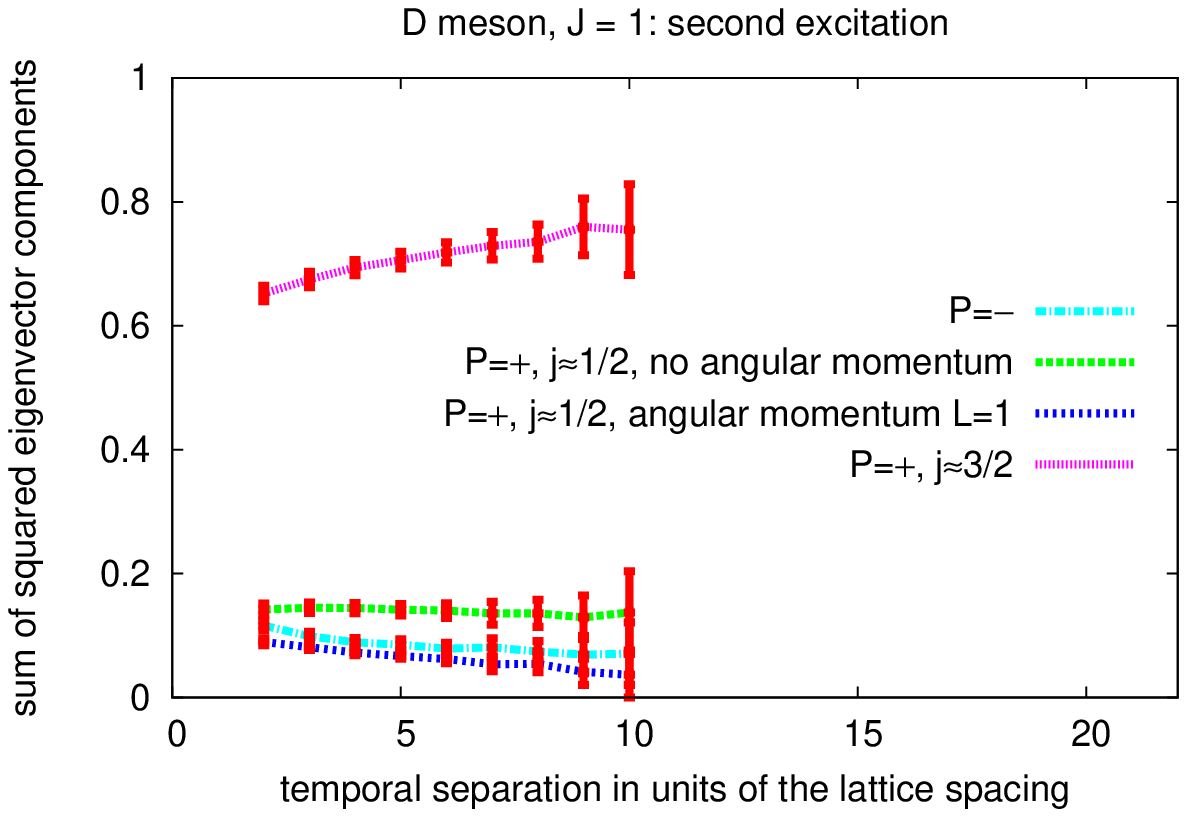} 
\end{center}
\vspace{-0.5cm}
\caption{\label{FIG003}The operator content of the three lightest $J = 1$ $D$ meson states. The ground state (upper plot) is identified as $D^*$ ($J^P = 1^-$), the first excitation (left plot) as $D_1(2430)$ ($J^P = 1^+$, $j \approx 1/2$), the second excitation (right plot) as $D_1(2420)$ ($J^P = 1^+$, $j \approx 3/2$).}
\end{figure}
\begin{itemize}
\item As expected the ground state is dominated by $P = -$ meson creation operators (the light blue curve corresponds to the sum of the squared eigenvector components of the six $P = -$ operators from (\ref{EQN101}) and (\ref{EQN102}), $\gamma_z [\gamma_0]$, $((\mathbf{n} \times \vec{\gamma})_z - \mathbf{n}_z \gamma_0 \gamma_5) \gamma_5 [\gamma_0]$ and $((\mathbf{n} \times \vec{\gamma})_z + 2 \mathbf{n}_z \gamma_0 \gamma_5) \gamma_5 [\gamma_0]$). This confirms that the ground state is the $D^*$ state ($J^P = 1^-$).

\item The first excitation is dominated by $P = +$ meson creation operators with $j \approx 1/2$. Operators without angular momentum ($\gamma_z \gamma_5 [\gamma_0]$; green curve) generate trial states with larger overlap than those with angular momentum $L = 1$ ($((\mathbf{n} \times \vec{\gamma})_z - \mathbf{n}_z \gamma_0 \gamma_5) [\gamma_0]$; dark blue curve). Consequently, the first excitation is identified as the broad $D_1(2430)$ state ($J^P = 1^+$, $j \approx 1/2$), where total angular momentum $J = 1$ is mainly realized by the quark spin and not by relative angular momentum $L = 1$ of the two quarks.

\item Finally the second excitation, which is close in mass to the first excitation, is dominated by $P = +$ meson creation operators with $j \approx 3/2$ ($((\mathbf{n} \times \vec{\gamma})_j + 2 \mathbf{n}_j \gamma_0 \gamma_5) [\gamma_0]$; magenta curve). Consequently, the second excitation is identified as the narrow $D_1(2420)$ state ($J^P = 1^+$, $j \approx 3/2$).
\end{itemize}

Note that one could consider even more meson creation operators, e.g.\ $j \approx 3/2$ operators with angular momentum $L = 2$.

An analogous analysis for $D_s$ mesons yields qualitatively identical results.


\section*{Acknowledgments}

M.K.\ and M.W.\ acknowledge support by the Emmy Noether Programme of the DFG (German Research Foundation), grant WA 3000/1-1, and by the Helmholtz Graduate School HGS-HIRe for FAIR. This work was supported in part by the Helmholtz International Center for FAIR within the framework of the LOEWE program launched by the State of Hesse.



\end{document}